\documentstyle[aps,12pt]{revtex}

\def\be{\begin{equation}}
\def\te{\end{equation}}
\def\bea{\begin{eqnarray}}

\def\tea{\end{eqnarray}}
\makeatletter 
\@addtoreset{equation}{section}
\makeatother  
       
\def\boxit#1{\vbox{\hrule height1pt
                   \hbox{\vrule width1pt\kern0.3cm
                         \vbox{\kern0.3cm
                               \hbox{$\displaystyle #1$}
                               \kern0.3cm
                              }
                         \kern0.3cm \vrule width1pt
                        }
                   \hrule height1pt
                  }
            }

\begin{document}
\title{Stochastic semiclassical cosmological models}
\author{{\footnotesize Esteban Calzetta$^{\dag }$, Antonio Campos$^{\ddag }$ and
Enric Verdaguer$^{\S }$}}
\address{$\dagger $Instituto de Astronom\'{\i}a y F\'{\i}sica del
Espacio (IAFE) and Departamento de F\'{\i}sica,\\ Universidad de
Buenos Aires,
Ciudad Universitaria,\\
1428 Buenos Aires, Argentina\\
$\ddagger $ Grup de F\'{\i}sica Te\`orica and Institut de F\'{\i}sica
d'Altes Energies (IFAE),\\ Universitat Aut\`onoma de
Barcelona, 08193 Bellaterra (Barcelona), Spain;\\
and Department of Physics, University of Maryland,\\
College Park, Maryland 20742, USA\\
\S\ Departament de F\'{\i}sica Fonamental and IFAE,
Universitat de Barcelona,\\ Av.
Diagonal 647, 08028 Barcelona, Spain}
\maketitle
\bigskip
\bigskip
\begin{abstract}
We consider the classical stochastic fluctuations of spacetime geometry
induced by quantum fluctuations of massless non-conformal
matter fields in the Early Universe. To
this end, we supplement the stress-energy tensor of these fields with a
stochastic part, which is computed along the lines of the
Feynman-Vernon and
Schwinger-Keldysh techniques; the Einstein equation is therefore upgraded
to a so called Einstein-Langevin equation. We consider in some detail
the conformal fluctuations of flat spacetime and the fluctuations
of the scale factor in a simple cosmological model
introduced by Hartle, which consists of a spatially flat
isotropic cosmology driven by radiation and dust.
\end{abstract}

\newpage 

\section{Introduction}

\label{sec:intro}When a model of the Early Universe is developed, it becomes
necessary to take into account the quantum nature of matter fields, even if
for a wide range of energies, leading up to Planck's, the geometry of the
Universe itself may be described in classical terms. The simplest and most
studied way to couple quantum matter to a classical gravitational field is
through the expectation value of the energy momentum tensor of the quantum
fields \cite{rosenfeld,BD82}; however, the quantum fields must
necessarily undergo fluctuations \cite{pazsin,hargell}, and it has
been recently shown that these fluctuations are by no means negligible in
many situations of interest \cite{fordkuo,hunich}.

To a certain extent, quantum fluctuations may be introduced in a classical
model as uncertainty in the initial conditions. However, fluctuations play a
subtler role when the semiclassical evolution, as it is in fact the rule, is
dissipative \cite{CH89,CH94}. In this case, the semiclassical
gravitational field interacts in a nontrivial way with the environment
provided by the quantum matter degrees of freedom \cite{Hu89}, leading, e.
g., to anisotropy damping and loss of information regarding initial
conditions \cite{FHH79,HH80,CH87,cal}. The environment
reacts back on the system in a way which is demanded by consistency of the
underlying quantum theory. Under equilibrium conditions in static space
times, the relationship between damping and back reaction becomes rather
simple, and it is embodied in the so - called fluctuation - dissipation
theorems\cite{fdt}. Similar, if more complex, connections between damping
and back reaction hold in nonequilibrium situations; as a rule, back
reaction acquires a stochastic character\cite{HS95HM95}.

In the particular case of semiclassical cosmology, we can be more precise
about what is going on. Except in some very special cases, such as
conformally flat universes coupled to conformal matter, or purely adiabatic
evolution, the cosmological evolution leads to particle creation \cite
{parker69,Zel70}. The phase relationships between the created particles are
quickly lost; even very weak interactions may accomplish this on very short
time scales\cite{HPZ93}. Under these conditions, particle creation is a
dissipative process. The back reaction of the matter fields will then
(loosely speaking) have a time reversible component associated to vacuum
polarization phenomena (such as trace anomaly) and a time oriented component
related to particle creation. The crucial point is that there is also a
third, stochastic component, induced by the quantum fluctuations of the
matter fields. To keep the average effect while disregarding these
fluctuations leads, as a rule, to an inconsistent theory.

In recent years Langevin type equations have been proposed to describe the
fluctuating back reaction of the matter fields on the classical
gravitational field, thus restoring the consistency of the semiclassical
Einstein equations. The new semiclassical Einstein-Langevin equations
predict classical stochastic fluctuations in the gravitational field \cite
{CH94,HS95HM95,CV96,ML96}. The derivation of these improved semiclassical
equations is based, usually, in the Closed Time Path (CTP) effective action
method and exploits the close connection that this effective action has with
the Influence Functional introduced by Feynman and Vernon \cite{FV63FH65}.
We should mention that a similar approach leading to Langevin type equations
has been used in recent years to study the generation of primordial
fluctuations in inflationary cosmologies \cite{CH95,sonia,andrew1,andrew2}
(Morikawa has independently arrived to similar conclusions regarding this
subject \cite{mor93}), as well as the more general problem of stochasticity
in effective field theory \cite{eft}.

In this paper we study the stochastic fluctuations arising from coupling
geometry to massless non-conformal fields in two cases of interest. One of
them is the study of the conformal fluctuations of flat spacetime. The
second case is a spatially flat Friedmann-Robertson-Walker (FRW) spacetime
with a classical source made of both radiation and dust. This model is
simple enough that the fluctuations can be studied in some detail, yet it
has significant features of a realistic cosmology; it was proposed some time
ago by Hartle \cite{Har81}, within the usual (Rosenfeld's) approach, to
study the back-reaction effects to the creation of massless non-conformal
particles in isotropic cosmologies. We shall not develop a full analysis,
but rather focus on the new features brought by the explicit introduction of
stochastic terms in the Einstein equations.

The Hartle model has two perfect fluids, one of them with the equation of
state of radiation and the other representing baryonic matter with the
equation of state of dust. A dimensionless parameter, $\xi $, measuring the
relative amounts of baryons and radiation, which is constant in classical
periods, is assumed to be always constant and with a value corresponding to
the present universe (which is very small, $\sim 10^{-27}$). A massless
non-conformal quantum field is coupled to this system. The presence of the
small portion of baryonic matter is essential to ensure particle creation,
since in a radiative FRW universe the scalar curvature vanishes and the
field cannot couple to the curvature. A perturbative expansion in terms of
the parameter $\xi $ is then seen to be equivalent to an expansion in a
parameter that measures the deviation from the conformal coupling. Since the
two degrees of freedom of the graviton field in a FRW background behave as
massless minimally coupled fields \cite{Gri75} Hartle's model provides a
good testing ground for the study of the back reaction due to the creation
of gravitons in cosmology.

The model describing the conformal fluctuations of flat spacetime due to
massless non-conformal fields can be easily obtained as a limit of Hartle's
model. One just needs to impose that there is no classical matter source and
that the cosmological scale factor is a small perturbation of unity. This
model, which is considerably simpler than the cosmological model, may be of
interest in connection with the semiclassical stability of flat spacetime 
\cite{Hor80} (see also Ref. \cite{FW96}) under conformal fluctuations.

We should stress that one of the main differences between the cosmological
back-reaction problem considered in this paper and other cosmological
back-reaction problems that have been considered in the literature is that
the later usually deal with conformal fields in isotropic cosmologies \cite
{parker69,Har77,FHH79}, or with conformal fields in cosmologies with small
anisotropies \cite{SZ72-77,BD80,Zel70,LS74LNSZ76HP78,HH80,CH87} or in
cosmologies with small inhomogeneities \cite{Fri89CV90CV92,CV94}. In the two
last cases particle creation takes place because the anisotropies or
inhomogeneities break the conformal invariance of the fields.

Cosmological models where the back-reaction equations have been solved
explicitly involve, usually, conformal fields in FRW cosmologies \cite
{Sta80,And83-84}. In such models there is no particle creation and all
quantum effects are a consequence of the fact that the vacuum expectation
value of the stress-energy tensor of the quantum field, subject to the trace
anomaly, does not vanish. Generally the stress-energy tensor is also subject
to a two parameter ambiguity connected to the addition of two conserved
terms quadratic in the spacetime curvature \cite{BD82,Wal94}. Such ambiguity
cannot be resolved within the semiclassical theory and depending on these
parameters one finds very different behaviors for the cosmological scale
factor. Different behaviors are also found as a consequence of the higher
than two time derivatives of the scale factor that appear in the
stress-energy tensor, this is a typical back-reaction phenomenon and the
origin of runaway solutions. In this connection several authors have argued 
\cite{Sim91PS93,FW96} that solutions to the semiclassical back-reaction
equations that deviate non perturbatively from the classical solutions
should be discarded.

When particle creation takes place, the back-reaction equations have been
solved only numerically for the case of massive conformally coupled fields
in FRW cosmologies \cite{And85-86}. In this case it is the mass that breaks
the conformal symmetry of the problem.

Here we are interested in the quantum effects due to the creation of
massless non-conformal particles. Our dynamical back-reaction equations for
the cosmological scale factor generalize that of Ref. \cite{Har81} in that a
new stochastic source is included. Because we start with a CTP functional
formalism our equations are real and causal and thus they admit an initial
value formulation. Thus when our equations are averaged with respect to the
stochastic source they should not be directly compared with the equations in
Ref. \cite{Har81}, which were obtained using an in-out functional formalism.
However the analysis carried out by Hartle on the behavior of the
cosmological scale factor at large and very small cosmological times is
still valid. For this reason we concentrate here mainly in the stochastic
effects which are new.

The plan of the paper is the following. In section II the actions for the
massless non-conformal scalar field, the gravitational field, and the
classical matter sources, which define Hartle's model, are described. In
section III we give the stochastic back-reaction equations for the
cosmological scale factor. The derivation of these equations is based in the
CTP effective action describing the interaction of the classical scale
factor with the quantum field, and the relation of this effective action
with the Influence Functional of Feynman and Vernon \cite{FV63FH65}. In
section IV we consider the particular case of the conformal perturbations of
flat spacetime. In section V we consider the fluctuation in the cosmological
scale factor in Hartle's model, and section VI summarizes our conclusions.

\section{Hartle's model}

In this section we describe a model \cite{Har81} in which a massless
non-conformal quantum scalar field is coupled to a spatially flat FRW
cosmology with a classical source made of radiation and dust. Since we use a
dimensional regularization technique some relevant terms in the classical
action are given in $n$ arbitrary dimensions. The cosmological model is
described by the $n$-dimensional spatially flat FRW metric, 
\begin{equation}
ds^2\ =\ a^2\eta _{\mu \nu }dx^\mu \ dx^\nu ,  \label{eq:metric}
\end{equation}
where $a(\eta )$ is the cosmological scale factor, $\eta $ is the conformal
time and $\eta _{\mu \nu }=diag(-1,1,\cdots ,1)$. The classical action for
the massless non-conformally coupled scalar field $\Phi (x^\mu )$ is

\[
S_m[g_{\mu \nu },\Phi ]=-{1\over 2}\int d^nx\ \sqrt{-g}\left[ g^{\mu \nu
}\partial _\mu \Phi \partial _\nu \Phi +\left( \xi _c+\nu \right) R\Phi
^2\right] , 
\]
where $R$ is the Ricci curvature scalar,
\[
R=2\left( n-1\right) \left[ \frac{\ddot a}{a^3}+\left( \frac{n-4}2\right) 
\frac{\dot a^2}{a^4}\right] ,
\]
$\xi _c+\nu $ is the parameter coupling the scalar field to the spacetime
curvature; $\xi _c\equiv {\frac{(n-2)}{4(n-1)}}$ is the coupling parameter
for a conformal field in $n$-dimensions. Since the metric (\ref{eq:metric})
is conformally flat it is convenient to introduce a new rescaled matter
field $\phi (x^\mu )$ by $\phi (x^\mu )\ \equiv \ a^{{\frac{n-2}2}}(\eta
)\Phi (x^\mu )$, and the scalar field action simplifies as, 
\begin{equation}
S_m[a,\phi ]\ =\ {\frac 12}\int d^nx\ \phi (x^\mu )\left[ \Box -\nu
a^2R\right] \phi (x^\mu ),  \label{eq:matter action}
\end{equation}
where now $\Box \equiv \eta ^{\mu \nu }\partial _\mu \partial _\nu $ is the
flat d'Alambertian operator. It is clear from this equation that if $\nu
\neq 0$ the conformally flat symmetry is broken by the coupling of the
scalar field.

The action for the gravitational field, the Einstein-Hilbert action $%
S_g^{EH} $, needs to be corrected with a counterterm to cancel the
divergencies that will come from the effective action. In our case it
suffices to add a term quadratic in the Ricci scalar. Note that terms
quadratic in $R_{\mu \nu }$ or the Weyl tensor could also be added but since
these terms are not necessary for the renormalization we simply assume that
their coefficients vanish. The divergent (in $n=4$) counterterm is: 
\begin{equation}
S_g^{div}[a;\mu _c]\ =\ \frac{\nu ^2\mu _c^{n-4}}{32\pi ^2\left( n-4\right) }%
\int d^nx\ \sqrt{-g}R^2,
\end{equation}
where $\mu _c$ is an arbitrary mass scale needed to give the correct
dimensions to this action term.

Since the only dynamical variable of the gravitational field is the
cosmological scale factor $a(\eta )$ the gravitational action reduces in our
case, when expanded in $(n-4)$, to 
\begin{equation}
S_g^{EH}[a]\ =\ {\frac{6{\cal V}}{\ell _p^2}}\int d\eta \ a\ddot a,
\label{eq:Einstein-Hilbert}
\end{equation}
\begin{eqnarray}
S_g^{div}[a;\mu _c] &&\ =\ {\frac{\nu ^2{\cal V}}{32\pi ^2}}\left[ {\frac{36%
}{n-4}}\int d\eta \ \left( {\frac{\ddot a}a}\right) ^2+36\ln \mu _c\int
d\eta \ \left( {\frac{\ddot a}a}\right) ^2\right.  \nonumber \\
\ \ &&\hskip1.5cm\left. +12\int d\eta \ \left\{ 3\left( {\frac{\ddot a}a}%
\right) ^2\ln a+\left( {\frac{\ddot a}a}\right) \left[ 3\left( {\frac{\dot a}%
a}\right) ^2+2\left( {\frac{\ddot a}a}\right) \right] \right\} \right] 
\nonumber \\
\ \ &&+O(n-4),  \label{eq:divergent action}
\end{eqnarray}
where $\ell_p^2=16\pi G_N$ ($G_N$ is the Newtonian gravitational constant)
and
${\cal V}$ is the volum integral ${\cal V}\equiv \int d^3{\bf x}$.

To the above scalar and gravitational actions we need to add the action for
the classical matter sources $S_m^{cl}$, which are a radiative and dust
perfect fluids. This action is given by, 
\begin{equation}
S_m^{cl}[a]\ =\ -{\cal V}\int d\eta \ \tilde \rho _ba,
\label{eq:classical action}
\end{equation}
where $\tilde \rho _b$ is a constant parameter related to the baryon energy
density. It is connected to a similar parameter $\tilde \rho _r$ for the
energy density of the radiation, through 
\begin{equation}
\xi =\ {\frac{\ell _p\tilde \rho _b}{\tilde \rho _r^{3/4}}},
\label{eq:rate parameter}
\end{equation}
which is of order $\xi \sim 10^{-27}$ for the present universe. This
parameter $\xi $ measures the baryon to photon ratio.

The action (\ref{eq:classical action}) is justified because when it is
varied with respect to the scale factor $a(\eta )$ it reproduces the trace
of the stress-energy tensor for a radiative perfect fluid with equation of
state $p_r=\rho _r/3$ ($p_r$ is the pressure and $\rho _r$ is the energy
density of the radiation), and of a dust like baryon fluid of pressure $%
p_b=0 $ and energy density $\rho _b$. When the dynamics of the scale factor
is driven by the classical source only one finds that $\rho _b=\tilde \rho
_ba^{-3}$ and $\rho _r=\tilde \rho _ra^{-4}$, then the relative amounts of
baryons and radiation as defined by $\ell _p\rho _b/\rho _r^{3/4}$ becomes
constant and given by (\ref{eq:rate
parameter}). In fact, when the
classical action (\ref{eq:classical
action}) is considered as the only
dynamical source of gravity from the Einstein-Hilbert action (\ref
{eq:Einstein-Hilbert}) one gets $6\ddot a=(\ell _p^2/2)\tilde \rho _b$ or
equivalently $R=-(\ell _p^2/2)T^{cl}$, where $T^{cl}=-\rho _b$ is the trace
of the stress-energy tensor of a perfect fluid of dust (baryons) plus
radiation.

We note that when baryons are not present, {\it i.e.} when $\xi =0$, the
scalar curvature vanishes ($R=0$), and from the action (\ref
{eq:matter
action}) for the scalar field $\phi (x^\mu )$ it is clear that
this field behaves as a free field in flat spacetime. Therefore with the
usual Minkowskian definition of a vacuum state (which corresponds to the
conformal vacuum for $\Phi (x^\mu )$) no particles can be created by the
expansion of the Universe. Back-reaction effects are still possible due to
the vacuum expectation value of the stress - energy tensor of the scalar
field, which in this case depends on the two free parameters mentioned in
the introduction. One of the parameters is associated to an action
proportional to $R^2$ and the other to an action proportional to $R_{\mu \nu
}R^{\mu \nu } $. The quadratic terms of the first parameter vanish when $R=0$%
, but the quadratic terms of the second parameter are generally different
from zero even when the scalar curvature vanishes. Thus all quantum effects
in this case depend crucially on the second parameter being different from
zero. We assume that such parameter is zero, thus $R=0$ is a consistent
solution to the semiclassical back-reaction problem. Since our interest is
in the back reaction due to particle creation we will take $\xi $ as a small
non zero parameter, and we will consider the $\nu a^2R$ term in (\ref
{eq:matter
action}) as a perturbative term which gives a measure of the
deviation from the radiative case $R=0$. In practice we will compute the
quantum corrections as perturbations in powers of the parameter $\nu $, and
this should be consistent with an expansion in $\xi $.

\section{Stochastic back-reaction equations}

\label{sec:stochastic equations}In this section we compute the stochastic
back-reaction equations for the cosmological scale factor $a(\eta )$ from
the CTP effective action to one-loop order for the quantum field $\Phi
(x^\mu )$ in the FRW spacetime (\ref{eq:metric}). This action is evaluated
using a perturbative expansion in powers of the parameter $\nu $. The CTP
effective action was introduced by Schwinger \cite{Sch61-62,Kel65,CSHY85};
for its application in a curved background see Refs. \cite
{CH87,CH89,Jor86-87,CV94}. The idea is to start with a generating functional
from which one obtains expectation values, instead of the matrix elements
one obtains using the generating functional of the ordinary in-out effective
action \cite{AL73}. The CTP effective action is the Legendre transform of
such generating functional. The price one pays for this is that one needs to
double the number of fields: the $(+)$ and $(-)$ fields below. For the
details of this paper we follow closely Ref. \cite{CV94}, see also Ref. \cite
{CV97}.

In our case we have two fields $a(\eta )$ and $\phi (x^\mu )$ whose free
actions are given by $S_g[a;\mu _c]+S_m^{cl}[a]$ from (\ref
{eq:Einstein-Hilbert}), (\ref{eq:divergent action}) and (\ref
{eq:classical
action}), and by $S_m^{free}[\phi ]={\frac 12}\int d^nx\ \phi
\Box \phi $ from (\ref{eq:matter action}), and an interaction action given
by $S_{int}[a,\phi ]=-{\frac \nu 2}\int d^nx\ a^2R\phi ^2$. The field $a$ is
classical, thus we only need to go up to one loop order for the field $\phi $%
, which corresponds to the first order expansion of the generating
functional in powers of $\hbar $. We have 
\begin{eqnarray}
\Gamma _{CTP}[a^{\pm },\bar \phi _{\pm }] &\simeq &S_g[a^{+};\mu
_c]-S_g[a^{-};\mu _c]+S_m^{cl}[a^{+}]-S_m^{cl}[a^{-}]  \nonumber \\
&&\ +S_m[a^{+},\bar \phi _{+}]-S_m[a^{-},\bar \phi _{-}]-{\frac i2}Tr(\ln G),
\label{eq:one-loop order}
\end{eqnarray}
where the fields $\bar \phi _{\pm }$ correspond to the expectation values of 
$\phi _{\pm }$, and $G$, which is a matrix operator with $2\times 2$
components, is the inverse of the classical kinetic operator $A=diag\left(
\Box -\nu (a^{+})^2R^{+},-(\Box -\nu (a^{-})^2R^{-})\right) $, obtained from
(\ref{eq:matter action}), with CTP boundary conditions (see \cite{CH87}).
Since we are only interested in the effective action for the scale factor $%
a(\eta )$, we set $\bar \phi _{\pm }=0$ and expand the functional
determinant in powers of the coupling constant; finally \cite{CV97}, the
renormalized CTP effective action reads: 
\begin{equation}
\Gamma _{CTP}[a^{\pm }]\simeq
S_{g,m}^R[a^{+}]-S_{g,m}^R[a^{-}]+S_{IF}^R[a^{\pm }],  \label{eq:CTP action}
\end{equation}
where 
\begin{equation}
S_{g,m}^R[a]\ =\ {\cal V}\int d\eta \ \left[ -{\frac 6{\ell _p^2}}\dot a^2+{%
\frac{9\nu ^2}{8\pi ^2}}\left( {\frac{\ddot a}a}\right) ^2\ln a-\tilde \rho
_ba\right] ,  \label{eq:renormalized classical action}
\end{equation}
and 
\begin{eqnarray}
S_{IF}^R[a^{\pm }] &&\ =\ {9\nu^2\over 8\pi^2}
\int d\eta \ d\eta ^{\prime }\ \Delta \left( {%
\frac{\ddot a}a}\right) (\eta ){\rm H}(\eta -\eta ^{\prime };\bar \mu
)\{\left( {\frac{\ddot a}a}\right) (\eta ^{\prime })\}  \nonumber \\
\ \ &&+36\nu ^2i\int_{-\infty }^\infty d\eta \int_{-\infty }^\eta d\eta
^{\prime }\ \Delta \left( {\frac{\ddot a}a}\right) (\eta ){\rm N}(\eta -\eta
^{\prime })\Delta \left( {\frac{\ddot a}a}\right) (\eta ^{\prime }),
\label{eq:influence action}
\end{eqnarray}
where we use the notation $\Delta f(\eta )\equiv f_{+}(\eta )-f_{-}(\eta )$, 
$\{f(\eta )\}\equiv f_{+}(\eta )+f_{-}(\eta )$, for an arbitrary function $%
f(\eta )$. The kernels $\mbox{\rm H}(\eta -\eta ^{\prime };\bar \mu )$ and $%
\mbox{\rm N}(\eta -\eta ^{\prime })$ are, 
\begin{eqnarray}
{\rm H}(\eta -\eta ^{\prime };\bar \mu ) &=&-\frac{{\cal V}}2\int \frac{dp^0%
}{2\pi }e^{-ip^0(\eta -\eta ^{\prime })} \ln \left[ -\left( {\frac{%
p^o+i\varepsilon }{\bar \mu }}\right) ^2\right]  ,  \nonumber \\
{\rm N}(\eta -\eta ^{\prime }) &&\ =\frac{{\cal V}}{32\pi }\delta (\eta
-\eta ^{\prime }).  \label{eq:kernels}
\end{eqnarray}
Note that the parameter $\mu _c$ has been absorved into the new parameter $%
\bar \mu $.

The subscript $IF$ in $S_{IF}^R[a^{\pm }]$ is used to indicate that this
part of the CTP effective
action is the influence functional action as defined by
Feynman and Vernon \cite{FV63FH65}. That is, it gives the effect of the {\sl %
environment}, the quantum field, on the {\sl system,} the cosmological scale
factor, which is the field of interest here. The identification between a
part of the CTP effective action and the influence functional was suggested
in Ref. \cite{CH94}.

\subsection{Improved effective action}

\label{subsec:effective action}The connection between the CTP effective
action in the semiclassical context and the influence functional action
introduced by Feynman and Vernon to describe the interaction between a {\sl %
system} and an {\sl environment} just mentioned gives an interesting new
light on the semiclassical back-reaction problem.

Let us now turn to the dynamical equations for the scalar field $a(\eta)$.
In principle these equations can be derived from the CTP effective action as 
$\left. \delta \Gamma_{CTP}/\delta a^+ \right|_{a^\pm=a} = 0$. A problem
might arise from the fact that such effective action has an imaginary part,
since $S^{R}_{IF}$ is complex, see (\ref{eq:influence action}). It should be
clear however that the imaginary part of $S^{R}_{IF}$ will not contribute to
the field equation derived in that form because $Im(S^{R}_{IF}[a^\pm])$ is
quadratic in the difference of the $(+)$ and $(-)$ fields and one finally
takes $a^+ = a^- = a$.

However, from the point of view of the {\sl system-environment}, relation
familiar in quantum statistical mechanics, the imaginary part of the
influence action is related to the noise suffered by the system from the
environment fluctuations \cite{FV63FH65}. This suggests an improvement to
the semiclassical back-reaction equations that takes into account such
fluctuations. This may be achieved if we define the influence functional 
\cite{FV63FH65} 
\begin{equation}
{\cal F}_{IF}[a^{\pm }]\ =\ e^{iS_{IF}^R[a^{\pm }]},
\label{eq:influence functional}
\end{equation}
and note that it may be written formally as 
\begin{equation}
{\cal F}_{IF}[a^{\pm }]\ =\ \int {\cal D}j\ {\cal P}[j]e^{i\left[
Re(S_{IF}^R[a^{\pm }])+6\nu \int d\eta \ j(\eta )\Delta \left( {\frac{\ddot a%
}a}\right) (\eta )\right] },  \label{eq:effective influence functional}
\end{equation}
with 
\begin{equation}
{\cal P}\left[ j\right] \ =\ \frac{e^{-{\frac 12}\int d\eta d\eta ^{\prime
}\ j(\eta )(\mbox{\scriptsize\rm N}(\eta -\eta ^{\prime }))^{-1}j(\eta
^{\prime })}}{\int {\cal D}j\ e^{-{\frac 12}\int d\eta d\eta ^{\prime }\
j(\eta )(\mbox{\scriptsize\rm N}(\eta -\eta ^{\prime }))^{-1}j(\eta ^{\prime
})}}.  \label{eq:functional distribution}
\end{equation}
That is, performing the path integral in (\ref
{eq:effective
influence
functional}) with ${\cal P}\left[ j\right] $
defined in (\ref{eq:functional
distribution}) leads directly to (\ref
{eq:influence
functional}). If we interpret ${\cal P}\left[ j\right] $ as a
Gaussian probability distribution, the action in (\ref
{eq:effective
influence
functional}) may be seen formally as the action
for a field $a(\eta )$ which is coupled to an external stochastic source $%
j(\eta )$.

Thus, the influence functional (\ref{eq:influence functional}) can be seen
as the mean value with respect to the stochastic field $j(\eta )$ of an
influence functional for an improved effective action $S_{eff}$ defined by, 
\begin{equation}
S_{eff}[a^{\pm };j]\ =\ S_{g,m}^R[a^{+}]-S_{g,m}^R[a^{-}]+Re(S_{IF}^R[a^{\pm
}])+6\nu \int d\eta \ j(\eta )\Delta \left( {\frac{\ddot a}a}\right) (\eta ).
\label{eq:effective action}
\end{equation}
This comes from (\ref{eq:effective influence functional}) and the addition
of the gravitational and classical matter terms (\ref
{eq:renormalized
classical action}). Now the field $j(\eta )$ will act as a
stochastic source in the improved semiclassical equation when the functional
derivation with respect to $a^{+}(\eta )$ is taken. This stochastic field is
not dynamical, it is completely determined by the following relations, which
may be derived from the characteristic functional, {\it i.e.} the functional
Fourier transform of ${\cal P}\left[ j\right] $, 
\begin{eqnarray}
\langle j(\eta )\rangle &&\ =\ 0 , \nonumber \\
\langle j(\eta )j(\eta ^{\prime })\rangle &&\ =\ {\rm N}(\eta -\eta ^{\prime
}).  \label{eq:correlations}
\end{eqnarray}
Since the probability distribution is Gaussian the noise kernel is the two
point correlation function of the stochastic field. In our case as one can
see from (\ref{eq:kernels}) the noise is white.

It is convenient for comparison with previous work \cite{Har81,And83-84}, to
write the effective action (\ref{eq:effective action}) in terms of
dimensionless quantities such as $\chi $, $b(\chi )$ and $\zeta (\chi )$ for
the conformal time $\eta $, cosmological scale factor $a(\eta )$ and
stochastic source $j(\eta )$, respectively. These are defined by 
\begin{equation}
\chi \ =\ {\frac{\tilde \rho _r^{1/4}}{6^{1/2}}}\eta ,\hskip1cmb\ =\ {\frac a%
{\ell _p\tilde \rho _r^{1/4}}},\hskip1cm\zeta \ =\ {\frac{\tilde \rho
_r^{1/4}}{6^{1/2}}}j.  \label{dlessvars}
\end{equation}

We also introduce dimensionless frequencies $\omega $, renormalization
parameter $\mu $ and volume $V$ instead of $p^o$, $\bar \mu $ and ${\cal V}$%
, respectively, as: 
\begin{equation}
\omega \ =\ {\frac{6^{1/2}}{\tilde \rho _r^{1/4}}}p^o,\hskip1cm\ \mu =\
6^{1/2}\ell _p\bar \mu ,
\hskip1cm{\cal \ }V\ =\ {\frac{\tilde \rho _r^{3/4}}{%
6^{1/2}}}{\cal V}.
\end{equation}
Then the improved effective action (\ref{eq:effective action}) for the
dimensionless scale field $b(\chi )$ becomes, 
\begin{eqnarray}
S_{eff}[b^{\pm };\zeta ] &\equiv &{\cal V}6^{1/2}\tilde \rho _r^{3/4}\int
d\chi \ \left\{ -\Delta (\dot b^2+ \xi b)(\chi )\right.  \nonumber \\
&&\ \hskip3.1cm+{\frac{\nu ^2}{32\pi ^2}}\left[ \Delta \left( \left( {\frac{%
\ddot b}b}\right) ^2\ln b\right) (\chi )-\Delta \left( {\frac{\ddot b}b}%
\right) (\chi )\kappa [\chi ;\left\{ {\frac{\ddot b}b}\right\} ]\right] 
\nonumber \\
&&\ \hskip3.1cm+{\frac \nu V}\zeta (\chi )\Delta \left( {\frac{\ddot b}b}%
\right) (\chi )\left. 
{ \atop }
\right\} .  \label{eq:final effective action}
\end{eqnarray}
The stochastic source $\zeta (\chi )$ is defined, after (\ref
{eq:correlations}), by $\langle \zeta (\chi )\rangle =0$ and 
\begin{equation}
\langle \zeta (\chi )\zeta (\chi ^{\prime })\rangle \ =\ {\frac V{192\pi }}%
\delta (\chi -\chi ^{\prime }).  \label{eq:noise}
\end{equation}
The non-local operator $\kappa $ is defined by its action on a test function 
$f(\chi )$ by, 
\begin{equation}
\kappa [\chi ;f(\chi )]\ \equiv \ \int_{-\infty }^\infty {\frac{d\omega }{%
2\pi }}\ e^{-i\omega \chi }h(\omega )f(\omega ),
\label{eq:non-local operator}
\end{equation}
where $f(\omega )$ is the Fourier transform of the test function, 
defined by $f(\omega)=\int_{-\infty}^{\infty} d\chi e^{i\omega\chi}
f(\chi)$,
and 
\begin{equation}
h(\omega )\ \equiv \frac 12\ \ln \left[ -\left( {\frac{\omega
+i\varepsilon }\mu }\right) ^2\right] ,  \label{eq:h kernel}
\end{equation}

This kernel is real and causal, as can be seen from its analytical
structure. Note that this is in fact the main difference between the CTP
approach and Hartle's approach.

\subsection{Stochastic back-reaction equations}

\label{subsec:equations}The dynamical equations for the cosmological scale
factor in its dimensionless form, $b(\chi )$, may be obtained by functional
derivation of the effective action (\ref{eq:final effective action}) as $%
\left. \frac{\delta \,\,\,}{\delta b^{+}}\left( S_{eff}[b^{\pm };j]\right)
\right| _{b^{\pm }=b}\ \equiv \ 0.$ After functional derivation and the
identification of the $(+)$ and $(-)$ fields, the equation acquire an
interesting form if we multiply it by $\dot b$, 
\begin{eqnarray}
&&\ {\frac d{d\chi }}\left\{ \dot b^2- \xi b-{\frac{\nu ^2}{32\pi ^2}}%
\left( {\frac{\ddot b}b}\right) ^2\ln b+\dot b{\frac d{d\chi }}\left[ {\frac{%
\nu ^2}{16\pi ^2}}\left[ {\frac{\ddot b}{b^2}}\ln b-{\frac 1b}\kappa [\chi ;{%
\frac{\ddot b}b}]\right] +{\frac \nu V}{\frac{\zeta (\chi )}b}\right]
\right\}  \nonumber \\
\ &&\hskip3cm={\frac{\ddot b}b}{\frac d{d\chi }}\left[ -{\frac{\nu ^2}{16\pi
^2}}\kappa [\chi ;{\frac{\ddot b}b}]+{\frac \nu V}\zeta (\chi )\right] .
\label{eq:stochastic semiclassical equation}
\end{eqnarray}

The effects of the quantum field are those proportional to the coupling
parameter $\nu $. The terms with $\ln b$ come from the renormalization of
the stress-energy tensor of the quantum field, see (\ref{eq:divergent action}%
), and there are non-local contributions in the terms with $\kappa [\chi ;%
\ddot b/b]$. The effects due to the creation of quantum particles, in
particular the dissipation of the field $b(\chi )$ are included in the
non-local term.

A point of interest is the role played by the parameter $\mu $ in the
equation (\ref{eq:stochastic semiclassical equation}).This parameter is
related to the local part of $\kappa [\chi ;f(\chi )]$, thus a change in
this parameter is equivalent to a change in the coefficients in some local
terms in the stress-energy momentum tensor of the quantum field. This
parameter in fact is related to the two parameter ambiguity in the
definition of such stress-energy tensor already mentioned. In our case one
of the parameters has been fixed by our renormalization procedure and the
ambiguity remains in the other parameter, see Ref. \cite{FW96} for more
details.

When quantum effects are ignored, {\it i.e.} when we take $\nu =0$, the
back-reaction equation becomes $2\ddot b-\xi =0$. This is the classical
equation when only the two classical fluids are present; it admits the
solution 
\begin{equation}
b\ =\ \chi +{\frac 14}\xi \chi ^2,
\end{equation}
with appropriate initial conditions. When baryonic matter is not present, 
{\it i.e.} $\xi =0$, the classical solution is $b=\chi $. If averaged with
respect to $\zeta (\chi ),$ eq. (\ref{eq:stochastic
semiclassical
equation}%
) also admits $b=\chi $ as a solution, which is in agreement with Hartle's
solution. As remarked in the introduction this is expected as in such case
no particle creation takes place. Had we introduced local terms in the
action of the type $R_{\mu \nu }R^{\mu \nu }$, this would not be necessarily
the case.

\section{Conformal fluctuations in flat spacetime}

\label{sec:flat space}The semiclassical back-reaction equation (\ref
{eq:stochastic semiclassical
equation}) was derived for a non-conformal
quantum field in a universe filled with classical radiation and baryonic
matter, using perturbation theory around a conformal vacuum state. The
classical contributions are in the first two terms of the equation {\it i.e.}
$\dot b^2-\xi b$. These two terms are obtained from the classical
stress-energy tensor and different classical sources will lead to different
terms. Thus by changing these first two terms in (\ref
{eq:stochastic
semiclassical equation}), the equation may be used in
different situations of interest. Besides the case of cosmological interest
described in the previous section, a case of obvious interest is that of the
conformal fluctuations of flat spacetime.

To obtain the equations for the fluctuations around flat spacetime, one
simple takes $\xi =0$ (no baryons) and notes that $b=1$ (or $a=1$) is the
conformal factor for flat spacetime. Thus if we take 
\begin{equation}
b(\chi )\ =\ 1+g(\chi )
\end{equation}
in (\ref{eq:stochastic semiclassical equation}) with $\xi =0$ and consider
the linear order in $g(\chi )$ we find an equation for the conformal
fluctuations of flat spacetime, interacting with a massless non-conformally
coupled scalar field. Equation (\ref{eq:stochastic
semiclassical equation})
in this linear approximation simplifies considerably, it can be integrated
twice and becomes 
\begin{equation}
g-{\frac{\nu ^2}{32\pi ^2}}\kappa [\chi ;\ddot g]+{\frac \nu {2V}}\zeta
(\chi )=A_o\chi +B_o,
\end{equation}
where $A_o$ and $B_o$ are integration constants, $B_o$ gives just a global
rescaling of the conformal factor and $A_o$ gives a linear expansion of the
scale factor. This linear term is pure gauge, it represents an infinitesimal
coordinate change of $t\rightarrow t+{\frac{A_o}2}t^2$, where $t$ is the
Minkowski time. We can take $A_o=B_o=0$ as our background spacetime or,
alternatively, define a new perturbation of the scale factor $G(\chi )\
\equiv \ g(\chi )-A_o\chi -B_o$, whose equation is 
\begin{equation}
G(\chi )-{\frac{\nu ^2}{32\pi ^2}}\kappa [\chi ;\ddot G]+{\frac \nu {2V}}%
\zeta (\chi )=0.  \label{eq:flat stochastic equation}
\end{equation}
The classical solution corresponds now to $G(\chi )=0$ (no perturbation).
The quantum effects here reduce to the two terms which depend on $\nu $ in (%
\ref{eq:flat stochastic equation}), the first, of order $\nu ^2$, is a
non-local term which is linear in $\ddot G(\chi )$, and the second is the
external stochastic source $\zeta (\chi )$. The first includes the effects
of particle creation, and the second drives the classical stochastic
fluctuations of the conformal factor. To solve this equation we may take its
Fourier transform,
\begin{equation}
G(\omega )\ =\ -{\frac \nu {2V}}\left[ {\frac{\zeta (\omega )}{1+{\frac{\nu
^2}{32\pi ^2}}\omega ^2h(\omega )}}\right] ,
\label{eq:Fourier transform}
\end{equation}
where ${\rm h}(\omega )$ is given in (\ref{eq:h kernel}). From eq. (\ref
{eq:noise}) we know that
\[
\left\langle \zeta \left( \omega \right) \zeta \left( \omega ^{\prime
}\right) \right\rangle =\frac V{96}\delta \left( \omega +\omega ^{\prime
}\right), 
\]
and so
\[
\left\langle G\left( \omega \right) G\left( \omega ^{\prime }\right)
\right\rangle =\frac{\nu ^2}{384V}\left| 1+\frac{\nu ^2\omega ^2}{32\pi ^2}%
h\left( \omega \right) \right| ^{-2}\delta \left( \omega +\omega ^{\prime
}\right), 
\]
which leads to the real time correlation
\begin{equation}
\left\langle G\left( \chi \right) G\left( \chi ^{\prime }\right)
\right\rangle =\frac{\nu ^2}{768\pi V}\int {d\omega\over 2\pi}
\;\frac{e^{-i\omega
\left( \chi -\chi ^{\prime }\right) }}{\left| 1+\frac{\nu ^2\omega ^2}{32\pi
^2}h\left( \omega \right) \right| ^2} .  \label{rtc}
\end{equation}

As $\nu \rightarrow 0$, the conformal stochastic fluctuations become white
noise, as it could be expected. Indeed, we could Taylor expand inside the
integration sign to develop the correlation function as a formal power
series in $\nu .$ An important problem, however, of this approximation is
that the fluctuation effect (in $\zeta (\chi )$) and the dissipation effect
(in $\kappa [\chi ;\ddot G]$) are treated at different orders of
approximation and this means that the fluctuation-dissipation relation
cannot be fully appreciated. In fact, one would expect that
as in the Brownian
motion the scale factor driven by the stochastic term will loose energy by
dissipation due to particle creation and that finally a sort of equilibrium
between the two effects will be reached. Here the stochastic effect
dominates the behavior of $b(\chi )$.

We may estimate the correlation time at finite $\nu $ by rescaling
$\omega =(8\pi /\nu)s $ , 
we find
\[
\left\langle G\left( \chi \right) G\left( \chi ^{\prime }\right)
\right\rangle =\frac \nu {96V}\int_{-\infty }^\infty 
{ds\over 2\pi}\;\frac{e^{-i8\pi \nu
^{-1}s\left( \chi -\chi ^{\prime }\right) }}{\left[ 1+s^2\ln \left( \frac{%
8\pi s}{\mu \nu }\right) ^2\right] ^2+\pi ^2s^4} . 
\]
The integral is controlled by the small $s$ range. Recalling that the
function $x\ln (x/e)$ has a minimum at $x=1$, we expand
\[
1+x\ln \left( \frac{\lambda x}e\right) \sim 1-\frac 1\lambda +\frac \lambda 2%
\left( x-\frac 1\lambda \right) ^2\sim 1+\frac \lambda 2\left( x-\frac 1%
\lambda \right) ^2 , 
\]
where
$
\lambda =e (8\pi /\mu\nu)^2
$, 
and we have used that $\lambda $ is a large number. In this approximation
\begin{eqnarray*}
\left[ 1+s^2\ln \left( \frac{8\pi s}{\mu \nu }\right) ^2\right] ^2+\pi ^2s^4
&\sim &1+\lambda \left( s^2-\frac 1\lambda \right) ^2+\pi ^2s^4 \\
&\sim &\left[ \lambda +\pi ^2\right] s^4-2s^2+1 .
\end{eqnarray*}
The denominator has poles at
\[
s^2=\frac 1{\lambda +\pi ^2}\left[ \pm i\sqrt{\lambda +\pi ^2-1}+1\right], 
\]
which converge towards
\[
s\sim \left[ \lambda +\pi ^2\right] ^{-1/4}\exp \left( 1+2k\right) 
\frac{i\pi }4 , 
\]
($k=0,1,2,3$) as $\lambda\rightarrow \infty $. Choosing the contour of
integration, we find
\begin{equation}
\left\langle G\left( \chi \right) G\left( \chi ^{\prime }\right)
\right\rangle \sim
{\nu^2\over 768\pi V} {\Theta\over\sqrt{2}}
\cos \left[ \Theta \left| \chi-\chi^{\prime}\right| -\frac \pi 4\right]
\exp \left[ -\Theta\left|\chi-\chi^{\prime}\right| \right],
\label{wkbcf}
\end{equation}
where
\[
\Theta =\frac{4\sqrt{2}\pi }{\nu \left[
\lambda +\pi ^2\right] ^{1/4}}. 
\]
We see that at finite $\nu $ the correlation decays exponentially with a
correlation time $\Theta^{-1} \sim \sqrt{\nu }$.

\section{Fluctuations in a universe with matter and radiation}

\label{sec:matter + radiation}We turn now to our main problem, which is the
analysis of the fluctuations of the cosmological scale factor in Hartle's
model, as described by (\ref{eq:stochastic semiclassical equation}).

The analytic solution of the stochastic integro-differential equation (\ref
{eq:stochastic semiclassical equation}) is not possible, but since $\xi $,
the parameter that gives the baryon to photon ratio is assumed to be very
small, it makes sense to find solutions linearized around a radiative
universe. Thus we follow Hartle \cite{Har81} and look for solutions of the
type 
\begin{equation}
b(\chi )\ =\ \chi +\xi g(\chi ).
\end{equation}
Substituting this into (\ref{eq:stochastic semiclassical equation}), taking
only terms linear in $\xi $, the equation can be integrated twice and
becomes 
\begin{equation}
g+{\frac{\nu ^2}{32\pi ^2\chi }}\left( {\frac{\ddot g}\chi }\ln \chi -\kappa
[\chi ;{\frac{\ddot g}\chi }]\right) +{{\frac \nu {2V\xi }}\frac \zeta \chi }%
\ =\ {\frac{\chi ^2}4}+A_o\chi +B_o,  \label{eq:perturbative equation (r+b)}
\end{equation}
where $A_o$ and $B_o$ are integration constants which may be taken as $%
A_o=B_o=0$, without loss of generality. The term $\chi ^2/4$ gives the
expansion corresponding to a matter dominated universe and comes from the
second term in (\ref{eq:stochastic semiclassical equation}). The terms with $%
\nu $ are of quantum origin. Equation (\ref{eq:perturbative equation (r+b)})
is the dynamical equation for the perturbation $g(\chi )$ of the scale
factor around a radiative classical solution.

This equation can now be directly compared with eq. (3.9) in Ref. \cite
{Har81}. In our case we have the external stochastic source $\zeta$ which
accounts for the fluctuations of the quantum stress-energy tensor and was
not considered previously, but if we take the mean value with respect of $%
\zeta$, the resulting equation is similar to Hartle's equation. The main
difference here, as we pointed out earlier, is that the equation is real and
causal. But the basic structure of the equation is similar and the
conclusions that Hartle's draws in his analysis apply also here.

In our case however we want to concentrate in the stochastic terms (which
are new) and the dissipative terms, and how they contribute to the dynamics
of $g(\chi )$. It is convenient to consider separately the behavior of
fluctuations far away from and close to the cosmological singularity.

\subsection{Fluctuations far from the singularity}

To study the behavior of the solutions to the Einstein - Langevin equation,
we first decompose $g$ into the deterministic part $g_d$ and the stochastic
component $g_s$. The behavior of the deterministic part follows Hartle's
original analysis \cite{Har81}, leading to $g_d\sim \chi ^2/4+0(\ln \chi
/\chi ^2)$ for large $\chi $.

As for the stochastic part, it is convenient to adopt as dynamical variable $%
f=\ddot g_s/\chi $, rather than $g_s$ itself. The variable $f$ satisfies
\begin{equation}
\int^\chi d\chi ^{\prime }\;\chi \chi ^{\prime }\left( \chi -\chi ^{\prime
}\right) f\left( \chi ^{\prime }\right) +{\frac{\nu ^2}{32\pi ^2}}\left\{
f\ln \chi -\kappa \left[ \chi ;f\right] \right\} =-{\frac \nu {2V\xi }\zeta }%
\ {,}  \label{feq}
\end{equation}

Instead of solving directly for $f$, it is convenient to look at the
propagator
\[
G(\chi ,\chi ^{\prime })=\frac{\delta f(\chi )}{\delta \zeta (\chi ^{\prime
})} 
\]
which obeys the equation
\begin{equation}
\int^\chi d\chi ^{\prime \prime }\;\chi \chi ^{\prime \prime }\left( \chi
-\chi ^{\prime \prime }\right) G(\chi ^{\prime \prime },\chi ^{\prime })+{%
\frac{\nu ^2}{32\pi ^2}}\left\{ \ln \chi G(\chi ,\chi ^{\prime })-\kappa
\left[ \chi ;G(\chi ,\chi ^{\prime })\right] \right\} =-{\frac \nu {2V\xi }}%
\delta (\chi -\chi ^{\prime })\ {,}  \label{Geq}
\end{equation}

The main formal difference between this case and the conformal fluctuations
of flat space time lies in that this equation is not time translation
invariant. However, as in the previous case we expect the propagator to
decay exponentially on the difference $u=\chi -\chi ^{\prime }$, with a
weaker dependence on some `center of mass' variable, which we take for
simplicity to be $X=(\chi +\chi ^{\prime })/2.$ Our problem is then to
disentangle the fast variable $u$ from the slow one $X$ \cite{kadbaym,CH88}.
Now observe that, since we expect $G(X,u)$ to be exponentially small unless $%
u\leq X$, we may approximate
\[
\ln \chi\,G(\chi,\chi^{\prime})\sim \ln X\, G(X,u). 
\]
On the other hand, let us write
\[
\kappa \left[ \chi ;G(\chi ,\chi ^{\prime })\right] \equiv \int d\chi
^{\prime \prime }\;h(\chi -\chi ^{\prime \prime })G(\chi ^{\prime \prime
},\chi ^{\prime }), 
\]
and set $\chi ^{\prime \prime }=\chi ^{\prime }+v$ to get
\[
\kappa \left[ \chi ;G(\chi ,\chi ^{\prime })\right] \equiv \int
dv\;h(u-v)G(X+\frac{v-u}2,v).
\]
Assuming that $G$ depends weakly on $X$, we obtain
\[
\kappa \left[ \chi ;G(\chi ,\chi ^{\prime })\right] \sim \int
dv\;h(u-v)G(X,v). 
\]
Handling the classical term in the same way,
\[
\int^\chi d\chi ^{\prime \prime }\;\chi \chi ^{\prime \prime }\left( \chi
-\chi ^{\prime \prime }\right) G(\chi ^{\prime \prime },\chi ^{\prime })\sim
X^2\int^udv\;(u-v)G(X,v), 
\]
we obtain the translation invariant equation
\[
X^2\int^udv\;(u-v)G(X,v)+{\frac{\nu ^2}{32\pi ^2}}\left\{ \ln
X\;G(X,u)-\kappa \left[ u;G\left( X,u\right) \right] \right\} =-{\frac \nu {%
2V\xi }}\delta (u), 
\]
where $X$ plays the role of a parameter. The solution reads
\[
G\left( X,u\right) ={\frac \nu {2V\xi }}\int \frac{d\omega }{2\pi }%
\;e^{-i\omega u}G(X,\omega ), 
\]
where
\begin{equation}
G(X,\omega )=\left\{ \frac{X^2}{\left( \omega +i\varepsilon \right) ^2}+{%
\frac{\nu ^2}{64\pi ^2}}\ln \left[ -\left( {\frac{\omega +i\varepsilon }{\mu
X}}\right) ^2\right] \right\} ^{-1} ,  \label{Gft}
\end{equation}
remarkably similar to its flat space counterpart. The self correlation is
given in terms of the propagator as
\[
\left\langle f\left( \chi \right) f\left( \chi ^{\prime }\right)
\right\rangle =\frac{ V}{192\pi }\int d\chi ^{\prime \prime }\;G(\chi
,\chi ^{\prime \prime })G(\chi ^{\prime },\chi ^{\prime \prime }), 
\]
which, with the same degree of accuracy, becomes
\[
\left\langle f\left( \chi \right) f\left( \chi ^{\prime }\right)
\right\rangle =\frac{{\cal \nu }^2}{768\pi V\xi ^2}\int
\frac{d\omega }{2\pi 
}\;e^{-i\omega (\chi-\chi^{\prime})}\left| G(X,\omega )\right| ^2 . 
\]
We can repeat for this self correlation the same analysis than in the
previous section, provided we replace $\left( \nu /X\right) $ for $\nu $ and 
$\mu X$ for $\mu $, to get
\begin{equation}
\left\langle g_s\left( \chi \right) g_s\left( \chi ^{\prime }\right)
\right\rangle \sim
{\nu^2\over 768\pi V \xi^2 X^2} {\Theta_c\over\sqrt{2}}
\cos \left[ \Theta_c\left|\chi-\chi^{\prime}\right| - {\pi\over 4}\right]
\exp \left[ -\Theta_c\left|\chi-\chi^{\prime}\right|\right] ,
\nonumber 
\end{equation}
where $\Theta_c=X\Theta$.
As a result, the correlation conformal time $\Theta_c^{-1}$ 
scales like $\sqrt{\nu }/X$. Of
course, the approximations we have made assume that this correlation
conformal time is small compared to $X$ itself ($X\geq \nu ^{1/4}$). In the
region of validity, we see that the effect of expansion is to shorten the
correlation time.

\subsection{Fluctuations near the singularity}

To obtain the form of the solutions valid at early times, we reason as
follows. When $\chi $ is small, the Fourier integral defining the nonlocal
part of the Einstein - Langevin equation is dominated by large frequencies $%
\omega \sim \chi ^{-1}$. But since $h\left( \omega \right) $ depends only
logarithmically on the frequency, we can then set
$
h\left( \omega \right)$ = constant = $h(\chi ^{-1})$, 
to obtain
\[
\kappa \left[ \chi ;f\left( \chi \right) \right] \sim -\ln \left[ \mu \chi
\right] f\left( \chi \right). 
\]
The Einstein-Langevin equation then becomes
\begin{equation}
\chi ^2g+{\frac{\nu ^2}{16\pi ^2}}\ln \left( \frac \chi {\chi _0}\right) 
\ddot g+{\frac{\nu \chi }{2V\xi }}\zeta \ =\ {\frac{\chi ^4}4}. \label
{eq:perturbative equation (r+b)1} 
\end{equation}
where $\chi _0=\mu ^{-1/2}$. Writing
$\chi =\chi _0e^{t^2/2} $,
we find
\[
\frac{\nu ^2}{32\pi ^2}\left[ \frac{d^2}{dt^2}-\left( \frac{1+t^2}t\right) 
\frac d{dt}\right] g+\chi _0^4e^{2t^2}g+\frac{\nu \chi _0^3}{2V\xi }%
e^{3t^2/2}\zeta =\frac{\chi _0^6}4e^{3t^2}. 
\]
Defining a new variable $v$ by 
$g=te^{t^2/4}v$,
we get
\[
v^{\prime \prime }+\frac 1tv^{\prime }+\left[ \frac{32\pi ^2
\chi _0^4e^{2t^2}
}{\nu ^2}-\frac 1{t^2}-\frac{t^2}4\right] v=
\frac{8\pi ^2\chi _0^6}{\nu ^2t}
e^{11t^2/4}-
\frac{16\pi ^2\chi _0^3}{\nu V\xi }e^{5t^2/4}\left( {\zeta\over t}
\right).
\]

Again we decompose $v$ into a deterministic and a stochastic part, $v_d$ and
$v_s$, respectively. The deterministic part admits a regular
solution as $t\rightarrow 0$
\[
v_d=\frac{4\pi ^2\chi _0^6}{\nu ^2}t\ln t+o\left( t^3\right). 
\]
In the same approximation, we find
\[
v_s=-\frac{8\pi ^2\chi _0^3}{\nu V\xi }\int_{t_1}^tdt^{\prime }\;
\left[ t-
\frac{t^{\prime 2}}t\right] \left( \frac \zeta {t^{\prime }}\right), 
\]
(more on $t_1$ later). Recalling that
\begin{equation}
\langle \zeta (t)\zeta (t^{\prime })\rangle \ =\ {\frac V{192\pi }}\delta
(\chi -\chi ^{\prime })={\frac V{192\pi }}\frac{e^{-t^2/2}}{\chi _0t}\delta
(t-t^{\prime }),  \label{whitecor}
\end{equation}
we get, always in the same approximation, that
\[
\left\langle v_s\left( t\right) v_s\left( t^{\prime }\right) \right\rangle
=\left( \frac{8\pi ^2\chi _0^3}{\nu V\xi }\right) ^2{\frac V{192\pi \chi
}}
_0\int_{t_1}^{t_{<}}\frac{ds}{s^3}\left[ t-\frac{s^2}t\right] \left[
t^{\prime }-\frac{s^2}{t^{\prime }}\right], \; 
\]
where $t_{<}=\min \left( t,t^{\prime }\right) $. Note that if we
assume that the stochastic source acts all the way from $t=0$, we get an
infinite amount of fluctuations.

As a matter of fact, we do not expect our equations to hold up to the
smallest times. First of all, our semiclassical universe presumably arises
out of a fully quantum one at the quantum to classical transition. The
semiclassical equation thus describes the model up to this time, but not
earlier. Besides, we do not expect the form (\ref{whitecor}) of the noise
correlation function to be true up to this time. The actual noise self
correlation, which ought to be derived from quantum gravity, presumably has
a finite correlation time (of the order of Planck's) and is not translation
invariant, but depends on the distance from its arguments to the `absolute
zero of time' (to borrrow Misner's phrase\cite{misner}). Eq.
(\ref{whitecor}
) is thus an approximation to be trusted only when the characteristic times
are much larger than Planck's. The lower cutoff $t_1$ can be thought of as
the earliest time when the model becomes meaningful.

While we cannot make a definite prediction for $t_1$, we can still observe
that if it is small enough, then fluctuations build up fast enough that they
may actually dominate the deterministic part over an stretch of time,
namely, when
\[
\left| \ln t\right| \leq \frac 1{\sqrt{6\pi }}\frac \nu {4\xi V^{1/2}\chi
_0^{7/2}t_1}. 
\]

It is interesting to look at the behavior of a quantity of direct physical
significance, such as the Hubble parameter ${\cal H}=\dot a/a^2$,
rather than the metric itself. In terms of dimensionless quantitities (cfr.
eq. (\ref{dlessvars})) ${\cal H}=H/\sqrt{6}l_P$, where
\[
H=\frac 1{b^2}\frac{db}{d\chi }. 
\]
Expanding $b(\chi )\ =\ \chi +\xi g(\chi )$ we can write $H$
in terms of $t$ ( $\chi =\chi _0e^{t^2/2}$)
and  $v$ as
\[
H=\frac 1{\chi ^2}\left[ 1+\frac \xi {\chi _0}e^{-t^2/4}\left(
\frac{dv}{dt}
+\left( \frac{2-3t^2}{2t}\right) v\right) \right]. 
\]
Now, decomposing as above $H=H_d+H_s$, we obtain 
\[
H_d\sim \frac 1{\chi ^2}\left[ 1+\frac{4\pi ^2\xi \chi _0^5}{\nu ^2}\left(
2\ln t+1\right) \right], 
\]
\[
H_s\sim -\frac{16\pi ^2\chi _0^2}{\nu V\chi ^2}\int_{t_1}^tdt^{\prime }\;\left( 
\frac \zeta {t^{\prime }}\right), 
\]
which leads to
\[
\left\langle H_s^2\right\rangle =\left( \frac{16\pi ^2\chi _0^2}{\nu V\chi ^2%
}\right) ^2{\frac V{384\pi \chi }}_0\left[ \frac 1{t_1^2}-\frac 1{t^2}
\right]. 
\]
We see that, provided $t_1$ is small enough, the fluctuations may have an
strong effect on the Hubble parameter.

\section{Conclusions}

In this paper we have considered the stochastic, classical metric
fluctuations induced by quantum fluctuations of matter fields in two cases
of interest, namely the conformal fluctuations of flat space time and a
simple cosmological model, the former being mostly of interest as
preparation for the latter. The quantum fluctuations act on geometry through
the energy momentum tensor, which has a deterministic part, associated to
vacuum polarization and particle creation, and also a fluctuating part,
related to the impredictable aspects of quantum behavior; the introduction
of the later leads to the formulation of the theory in terms of a so-called
Einstein-Langevin equation.

While the physical rationale for including such terms is clear (and earlier
estimates point to their quantitative relevance \cite{fordkuo,hunich}) and
the technical means to introduce them consistently into Einstein equations
are by now well established \cite{CH94,CH95,HS95HM95,CV96,ML96}
this is, to our
knowledge, the first time actual non trivial solutions to the 
Einstein-Langevin equation are obtained. 
The reason why the solution to particular
models lagged so much behind the formulation of the basic principles lies of
course in the staggering complexity of Einstein theory, involving issues of
gauge invariance, renormalization and nonlinear dynamics. The model proposed
by Hartle in Ref. \cite{Har81} is an ideal testing ground because it retains
most of the physical features of a realistic cosmological model, while being
simple enough to be tractable.

Since in this last analysis massless non-conformal fields in FRW
backgrounds are
of interest mostly as a toy model for gravitons \cite{Gri75} (for the
application of similar ideas to inflationary cosmology, see \cite
{mor93,CH95,sonia,andrew1,andrew2}), the effects we have discussed here are
at bottom quantum gravitational, and they are most relevant in the earliest
stages of cosmic evolution. This adds a degree of uncertainty in the theory,
associated to the present lack of understanding of the details of the
quantum to classical transition; the best we can do is to localize our
ignorance in a few undetermined parameters, such as the time at which the
initial conditions for semiclassical evolution are set and the value of the
several renormalized parameters in the model (in particular, we have applied
Occam's razor in setting to zero any parameter not specifically demanded by
renormalization).

Given these limitations, the main conclusions of our effort are that the
Einstein-Langevin equation may indeed be solved, that a consistent picture
of semiclassical evolution emerges and, most important, that this picture is
significatively different from what earlier (non stochastic) semiclassical
models have led us to believe (consider for example the ratio of the
fluctuating to the deterministic parts of the Hubble parameter, as discussed
in last section). Clearly more work shall be needed before questions of
direct cosmological impact (such as whether stochastic fluctuations may be
instrumental in creating the homogeneous patches where inflation becomes a
possibility, or rather work against the stability of those patches) may be
addressed. We continue our research on this, in our opinion, most relevant
issue of pre-inflation cosmology.

\begin{center}
{\normalsize {\bf ACKNOWLEDGMENTS}}
\end{center}

We are grateful to Bei-Lok Hu, Rosario Mart{\'\i }n, Diego Mazzitelli, Juan
Pablo Paz and Josep Porr{\`a} for very helpful suggestions and discussions.
This work has been partially supported by the European project CI1-CT94-0004
and by the CICYT contrats AEN95-0590 and AEN95-0882, Universidad de Buenos
Aires, CONICET and Fundaci\'on Antorchas.

\end{document}